# Moving from the Household to the Individual: Multidimensional Poverty Analysis

Ramya Vijaya
*Richard Stockton College*
*New Jersey, USA*
*Ramya.Vijaya@stockton.edu*

Rahul Lahoti
Indian Institute of Management Bangalore
Bangalore, India
*rahul.lahoti@gmail.com*

Hema Swaminathan
Indian Institute of Management Bangalore
Bangalore, India
*hema.swaminathan@iimb.ernet.in*

**Abstract**
Current multidimensional measures of poverty continue to follow the traditional income poverty approach of using household rather than the individual as the unit of analysis. Household level measures are gender blind since they ignore intra-household differences in resource allocation which have been shown to differ along gender lines. In this study we use new data from the Karnataka Household Asset Survey (KHAS) to construct an individual level multidimensional poverty measure for Karnataka, India. Our results show that an individual level measure can identify substantial gender differences in poverty that are masked at the household level. We also find a large potential for misclassification of poor individuals as non-poor when poverty is not assessed at the individual level.

## 1. Introduction

The dissatisfactions with purely monetary approaches to poverty measurement are by now well established. These led to the development of alternative measures that acknowledge that the experience of poverty is more complex and nuanced than merely a shortfall in income and consumption. Multidimensional measures of poverty, thus, conceptualize poverty along a spectrum of deprivations encompassing various aspects of wellbeing such as economic, social and material. However current multidimensional measures continue to mirror an important limitation of the traditional income approach. Income poverty is derived from household aggregate incomes even though poverty numbers are always referenced with respect to individuals. From a gender perspective, equating the household with the individual is particularly problematic as gender is an important axis of differentiation with men and boys often privileged over women and girls. This critique is also applicable to multidimensional poverty measures. Largely due to the lack of sex-disaggregated data, current multidimensional poverty measures continue to use the household as a unit of analysis. Thus, while the multidimensional measures help in unpacking the range of deprivations faced by a household, they are silent on individual experiences of poverty and remain gender blind.

Both theoretical and empirical literature are in agreement that not all individuals within a household are equal (Agarwal, 1997; Doss, 2005; Duflo, 2003; Quisumbing et al. 1995). Such differences are rendered invisible when poverty and deprivation are defined by household averages. To the extent that poverty analysis has addressed gender inequalities it has been through the feminisation of poverty narrative of the 1990s (Chant, 2012) and has been confined to investigating differences between male-headed and female-headed households. The implicit assumption here is that analysis by the sex of the head is a proxy for gender analysis (Lampietti and Stalker, 2000). This approach is highly inadequate with a key criticism being that it completely ignores women in male-headed households.

Incorporating a perspective on how poverty may be experienced by household members can aid policy makers in the design and evaluation of anti-poverty and livelihoods creation programmes. Since individuals within households can experience different kinds of deprivations, a household level multidimensional analysis does not give enough information about the interventions that might be most suitable for individuals based on gender, age etc. More importantly, a household level analysis does not allow an identification of individuals, both men and women, who might be experiencing severe deprivations even within 'non-poor' households.

Using data from Karnataka Household Asset Survey (KHAS), we construct a multidimensional measure of poverty separately for individuals within households and use a gender lens to sift through the results. KHAS collected data at individual level on several dimensions, giving us the opportunity to calculate an individual multi-dimensional poverty measure. To the best of our knowledge, this is the first study to generate an individual level multidimensional poverty measure. By comparing household and individual poverty estimates, we demonstrate that valuable information on who is poor is gained when the individual rather than the household is used as the unit.

The results indicate that a majority of individuals in households classified as non-poor are classified as poor in the individual analysis. Also gender differentials in poverty which are almost non-existent in household analysis become prominent in individual analysis. The difference between women and men, who are classified as poor, increases from 1 percentage point in household analysis to 39 percentage points in individual analysis. Even using the sex of the head of household in household level analysis masks these gender differences in poverty. These results point to large errors in classification of individuals as non-poor when using household level analysis. Finally we find that the experience of poverty is not the same for everyone. We are able to identify important differences in the kinds of deprivations that contribute to the poverty of different groups. These differences, which do not come to light at the household level, have important policy implications.

In the following section we provide a background to multidimensional poverty and discuss the engendering of poverty analysis. The rest of the paper is organized as follows. Section 3 presents the components of our multidimensional poverty measure. Section 4 describes the data and the methodology. Section 5 presents the results and section 6 concludes.

## 2. Engendering Poverty Measurements

Based on Amartya Sen's capabilities approach to human development, multidimensional poverty analysis attempts to extend the measurement of poverty to the functioning and capabilities space from the income and expenditure space. Both income and expenditures based measures have faced much criticism and controversy for being an arbitrary set of numbers that do not give a real sense of the deprivations facing the poor (Pogge and Reddy, 2010).

Beyond measurement issues, purely money-based poverty lines have a key conceptual problem highlighted by Amartya Sen in his capabilities approach. Income or money represents the means to better living conditions but it is not the better living condition in itself. A movement out of poverty should represent a reduction in deprivations and an actual improvement in living conditions or functionings that people can achieve. While income represents the ability to purchase commodities that help achieve some functionings, the conversion of commodities into functionings is not precise. Individuals differ in their ability to convert commodities into functionings due to various factors such as age, gender or physical ability. Age, gender and status within households can also impact the way commodities are distributed within households. Commodities often have to be supported by public goods (for example having access to schooling is necessary to achieve education) in order to achieve the functioning (Alkire, 2002; Sen, 1999). Moreover not all functionings are derived from commodities. For instance, one of the important basic human functionings in Sen's approach is the freedom to choose or exercise one's individual agency (Sen, 1988). Even for the wealthy, individual agency is often circumscribed by gender, age, marital status etc.

However, current multidimensional measures fall short of being able to fully integrate the gender dimensions of poverty. Typically, multidimensional measures have mirrored Gross Domestic Product (GDP) estimates in that they are based on country averages. One of the earliest efforts was the Human Development Index (HDI) pioneered by the United Nations Development Programme (UNDP). The HDI focused on longevity, educational attainment and standard of living. The Gender Development Index (GDI) and the Gender Empowerment Measure (GEM) attempted to bring in a gender perspective, though they did not include any specific gendered dimension of poverty such as time use, exposure to violence and so on (Bessell, 2010). It is also argued that by presenting average achievement figures for the country as a whole, such indices divert the focus from the poor (Pogge, 2010).

More recently, the availability of better household data has allowed multidimensional methods to focus specifically on deprivations among poor households. The most ambitious effort to implement a multidimensional measure of poverty has been the Multidimensional Poverty Index (MPI) introduced in the 2010 Human Development Report (Alkire and Santos, 2010). The MPI evaluates poverty based on a household's deprivation in three basic dimensions – education, health and living standards. Various indicators are used to measure each of the dimensions and they represent a mix of commodities and actual functionings. The three dimensions are equally weighted and a household's total deprivation score is compared to an established poverty cutoff. Since the MPI focuses on information from each household, as opposed to country averages, it is possible to consider the multiple and interconnected deprivations for the household, enabling identification of not only the poverty headcount ratio but also the intensity of poverty.

Alkire and Seth (2008) also developed a separate household Index of Deprivation for India using the third wave of the National Family and Health Survey (NFHS-3) data. Identification of the poor in India, referred to as 'below poverty line' (BPL) households, since 2002, explicitly acknowledges that poverty has multiple aspects. Thirteen socio-economic parameters including size of land, type of house, food security, clothing, sanitation, literacy, means of livelihood and indebtedness were used to identify whether a household qualified for the BPL status. This method comes closest to multidimensional poverty measurement techniques but was plagued with conceptual and data quality issues. Alkire and Seth (2008) contend that identifying the poor using their multi-dimensional Index of Deprivation, is both efficient and provides greater insight into dimensions of poverty across the various states.

However the Index of Deprivation for India and the MPI are both once again based on household level information. They do not shed any light on poverty as experienced by individuals since the household is treated as the unit of analysis. This is largely driven by data considerations as individual level information is not available for the some of the dimensions. Alkire and Santos (2010) while acknowledging this issue, argue that using household information has certain advantages, particularly for goods that are semi-public in nature and shared across household members. The lack of individual level analysis ignores important consideration of intra-household dynamics and inequalities within the household and therefore these measures continue to be gender blind.

Analysis of gender issues in poverty is limited to the feminisation of poverty narrative where the hypothesis was that women-headed households were increasingly becoming poorer than other households.[1] This claim, however, has not been validated empirically. Using longitudinal data Medeiros and Costa (2008) conclude that feminisation of poverty has not occurred in the eight Latin American countries they studied. Their findings are invariant to different measures and definitions of poverty. Overall, the research findings from developed and developing countries do not warrant the acceptance of the feminisation of poverty narrative as a stylized fact (see Medeiros and Costa (2008) for additional references). The few studies that exist for Indian data also do not find women-headed households as being systematically poorer than male-headed ones. Based on nationally representative data for India, Drèze and Srinivasan (1997) report that for rural India, women-headed households are actually less poor. A more recent study suggests that in addition to sex of the household head, marital status is an important consideration when discussing poverty incidence (Gangopadhyay and Wadhwa, 2004). Their results show that not currently married women-headed households are more susceptibility to poverty, which the authors conclude is due to their lower educational levels in comparison to their male counterparts.

Additionally, there are other concerns with the feminisation of poverty theme. Poverty was conceptualized in the traditional manner of either income or consumption shortfalls. This has been criticized for its narrow focus on money metric measures while ignoring other domains where deprivations may be experienced and are of particular significance to women. Typically, women have lower achievements in health, education, nutrition, decision making powers etc. Another consideration pertinent to women is that income based measures fail to distinguish between the availability of income and the actual control of and disbursement of the income among members of the household (Bessell, 2010).

Equally important, as Sen (2010) argues, the emphasis prioritizes a household-level focus rather than a real consideration of intra-household dynamics. It also reduces gender analysis to a narrow focus of women in poverty. Gender analysis is necessarily more expansive and should include a view of how poverty as a gendered process affects all households and its members, not just women or women-headed or poor households. In fact, analysis by sex of the household head can present an inaccurate picture of poverty. Diana Deere, Alvarado, and Twyman (2012) using data from Latin America and Caribbean show that for certain categories of assets, gender inequality is overestimated as headship-based analysis ignores women in male-headed households. Further, the use of headship analysis serves to homogenize all women within these two categories. There is little or no unpacking of women by other parameters that are also gendered in nature such as age, marital status, caste, and religion.

Ultimately, genuine gender analysis of the kind that moves beyond headship status needs sex-disaggregated data which is often not readily available. In this paper we are able use sex-disaggregated data from the KHAS dataset.

## 3. Dimensions of Poverty

The multidimensional poverty measure developed in this paper includes four dimensions; education, living standards, ownership of productive assets and empowerment. There is universal acceptance about the relevance of education and basic standard of living indicators in categorizing and understanding poverty. These indicators are part of the Millennium Development Goals (MDG) adopted by the United Nations. These attributes if inaccessible to households and individuals can have a profound impact on their current and future wellbeing. Since these dimensions have been extensively discussed in the literature (Alkire and Santos, 2010; Sachs 2005), here we focus on the latter two dimensions.

A notable omission in this study is that of the health dimension. It was not possible to include a health dimension as the KHAS data does not have the relevant information. The well documented persistence of gender disparities in health status imply that the deprivation gap between men and women reported here is likely an underestimate of the true gap underscoring the point that household poverty experiences cloak inequalities in individual differences.

In moving from the household to addressing individual deprivations, there are both conceptual and empirical challenges. Certain household dimensions are semi-public in nature, that is, they are non-excludable. For example, availability of a toilet can be classified as a semi-public good that all individuals within the household can use and derive well-being from. Moreover there is no empirical way to determine specific individual ownership of the toilet. In such cases the households and the individual's deprivations (or lack of deprivation) would be identical.

*Productive assets dimension:*

The stock of assets possessed by a household permits a longer-term perspective of economic security in way that is not possible using income or consumption data. Asset portfolios reflect both past and future income-generation opportunities through their contribution to livelihood choices, and the potential for participating in financial markets, generating rents, interests on savings, and profits from business. The characteristics of assets can impact the experience of poverty in several ways. Assets also provide a safety net during times of economic crises, through their sale or pawning to cope with an income shortfall. There is growing recognition that the composition of the asset basket can be a powerful force in mediating the experience of poverty. Households with few or no productive assets are typically more vulnerable to long term or chronic poverty than households that possess some level of these assets but experience income fluctuations (Carter and Barrett, 2006).

A gendered analysis of asset ownership and its implications for poverty are largely limited to the headship concept. An exception is provided by the bargaining literature that finds women's ownership of assets exerts a positive influence on their ability to participate in household decision making (Allendorf, 2007; Garikipati, 2009; Swaminathan, Lahoti, and J Y, 2012) while reducing their vulnerability to violence (Bhattacharyya, Bedi, and Chhachhi, 2011; Friedemann-Sánchez, 2006; Panda and Agarwal, 2005). Further, it is also beneficial for children's schooling and health outcomes (Allendorf, 2007; Katz and Chamorro, 2003).

It can be argued that asset ownership does not specifically describe an individual or a household's current state of living as some of the living standards indicators or education indicators do. In essence, it does not constitute living standards but is a determinant of living standards and is inappropriate to combine an asset dimension with the other constituent dimensions in evaluating multidimensional deprivation. However, we take the stand that the security of having a buffer against shocks in itself can create a sense of wellbeing and control over life circumstances and is a critical ingredient for improving one's quality of life. Further, the inclusion of ownership of productive assets criteria is a well-established part of the discourse regarding poverty measurement in India. Size of land and the type of housing have been included in the identification criteria for the Below Poverty Line census since 1997 (BPL). While there are several critiques of the BPL methodology, alternate methods suggested continue to acknowledge the relevance of ownership of land and house for poverty analysis in the India context (Dreze and Khera, 2010).

We recognize that ownership can be divorced from control over these resources, particularly for women. However, current data constraints (as described below) do not allow the use of transaction rights (ability to sell, rent, collateralize or bequeath) to be incorporated in the individual level analysis.

*Empowerment dimension:*

The concept of empowerment is complex, hard to define and does not lend itself to an easy set of metrics (Malhotra, Schuler and Boender 2002). Kabeer (2001) discusses empowerment in terms of the ability to make choices that could have an important impact on one's life. Ownership of assets is sometimes also considered a proxy for empowerment (Garikipati 2009). However, this is problematic as it equates resources as being both necessary and sufficient condition for empowerment. Malhotra, Schuler and Boender (2002) argue that while resources can act as a catalyst, women's agency is required to effectively utilise the resources for advancing their goals and interests. This paper uses mobility indicators which provide an insight into women's freedom of movement outside their household and community. These indicators are culturally specific and are often used as a proxy for empowerment in the South Asian context (Hashemi et. al 1996; Jejeebhoy 2000; Alkire and Seth, 2008). They relate to women's mobility and their ability to travel independently to places outside their home. We also include an additional question on women's ability to access health care services for themselves as it is a decision that has direct implications for their well being.

4. **A Multidimensional Poverty Measure for Karnataka**

This paper is based on the Karnataka Household Asset Survey (KHAS) 2010, collected by the Indian Institute of Management Bangalore, as part of a larger research project aimed at assessing gender and intra-household disparities in asset ownership. KHAS is a state-representative survey of Karnataka State, located in southwest India. Karnataka can be categorized into four agro-climatic zones; the northern and southern plateaus, the coastal areas and the mountainous Western Ghats. A stratified random sampling method was used with the survey covering eight districts across the four agro-ecological zones. The final sample comprising 4,088 households is representative of both rural and urban areas (see Swaminathan, J Y, and Lahoti, 2011 for further details on sampling).

A household and an individual questionnaire were administered by KHAS. The household questionnaire, in addition to the standard socio-demographic information, included an asset inventory to capture ownership details, mode of acquisition, and valuation data. In addition to the convention of recording if the household owned an asset or not, the IDs of all the owners were recorded on the questionnaire which permits an individual level analysis of asset distribution.

The individual questionnaire, among others, obtained information on transaction rights over assets (if the respondent was an asset owner), financial assets, and decision making processes within the household. Both household and individual questionnaires were administered to a primary respondent defined as the household member who had most knowledge of the economic circumstances of the household. If the primary was married, then his or her spouse was interviewed as the second respondent. If the primary was not married or the spouse was not available, a second respondent was chosen according to a predetermined set of procedures. Asset ownership details were verified with the second respondent as well. A total of 7,185 individuals from 4,088 households were interviewed for the survey.

Using the four dimensions – education, living standards, ownership of productive assets and empowerment – this study develops a household poverty measure as well as an individual poverty measure for all adults (18 years and older). Table 1 presents the indicators chosen to represent the dimensions and the cuts off we establish for each indicator. While the dimensions remain the same for the individual and the household level poverty measure, a few indicators are varied to capture intra-household differences. In order to ensure comparability, the individual and the household measures are calculated for the same sample of households.

**Table 1. KHAS-MPI dimensions**

| Dimension | Indicator | Deprivation | |
| --- | --- | --- | --- |
| | | Household | Individual |
| Education | Literacy | No adult member has completed at least primary education, i.e., 5 years of schooling | If he/she has not completed at least primary education |
| | Child Enrollment | A child in the age group 5-9 is not enrolled in school | Not included |
| Living standards | Electricity | No electricity | Same as household deprivation for all living standard indicators |
| | Floor | Floor is earth/mud | |
| | Sanitation | No toilet or has to share a toilet | |
| | Water | Water is not from piped source, borewell or closed/open well | |
| | Cooking fuel | Cooking fuel is *not* Electricity, LPG or Biogas (it is wood, charcoal, dung etc.) | |
| | Consumer durables | Owns less than two of either fan, TV, cell phone, cycle, refrigerator and two-wheeler; and does not | |
| | | | own a car or other four-wheeler |
| Productive assets | Primary residence Agricultural land | Does not own at least one of the two assets- agricultural land or primary residence | Household does not own at least one of the assets Individual does not own (individually or jointly) at least one of the assets |
| Empowerment | Allowed to travel to <br> 1. Market <br> 2. Health facility <br> 3. Natal home <br> 4. Outside village / community/area <br> 5. Decision to access health services for own needs | Assigned value of the women members | 1-4: Not allowed to travel alone <br> 5: Decision made by women with permission or by someone else <br><br> All females in household are attributed the deprivation score of the female respondent. Men are assumed to be non-deprived |

*Education:* The first indicator is based on the idea of proximate literacy, discussed by Basu and Foster, (1998) where the presence of one literate individual provides positive externalities for the entire household. Thus, a household with one literate member is better off in comparison to households with no literate members. This concept does not extend to the individual level as they may have differing access to the literate member. Moreover, differences in literacy between members of a household could impact the power dynamics within the household. It is conceivable that the bargaining capacities of an illiterate husband and illiterate wife might be more equal than those of a literate husband and an illiterate wife with consequences for household dynamics and resource allocation. Therefore, in the individual measure of multidimensional poverty, we consider only the individual's own level of education.

In the household level analysis we also include the child enrollment indicator. Universal primary education is recognized as one of the key components of the MDGs

Moreover this indicator gives us an idea about the generational trend in education. If a household currently has no adult literate member but the school age children are enrolled in primary education, there is the potential for future literacy for the household. In addition, enrollment rates in Karnataka (ASER 2011) have reached well above the 90 per cent level in recent years. Given this trend, the inability to send all primary school age children to school can indicate acute distress and poverty for a household. If a household has no children in that age cohort, however, they are non-deprived in this indicator.

*Living Standards:* We follow Alkire and Santos (2010) in deriving the deprivation cut offs for the living standard indicators. In the measure of individual poverty, we treat living standards as a public good accessible equally by all individuals in the households. Each individual was therefore assigned the values of their household's living standards indicators in our individual poverty measure. If a household has electricity, access to sanitation and proper flooring, the benefits would automatically be available to all members. Indeed it would be difficult to establish how an individual member might be excluded from the benefit of these amenities. A similar approach is adopted for consumer durables even though differential access is possible. However this is hard to measure, since there are no clear ownership documents for most consumer durables. In fact, most households considered consumer durables to be jointly owned by all members of the household.

Given the gendered nature of roles and responsibilities, lack of clean cooking fuel and access to safe drinking will likely impact women more than men. Recent evidence shows that indoor air pollution from contaminated cooking fuel has a disproportionately large impact on women's respiratory health (Duflo, Greenstone, and Hanna, 2008). Similarly, the lack of a dependable water source or access to piped water will certainly increase women's work burden while also contributing to time poverty. However, in the absence of a detailed time use module it is impossible to pinpoint which woman in the household is most affected. It can be argued that deprivation in water and cooking fuel will have some secondary impact on all members even if they do not directly participate in the activity. This can take the form of some degree of air pollution from cooking fuel and an experience of water shortage for personal use.

*Productive Assets:* Productive asset dimension at the household level is evaluated on the basis of a household's ownership of assets. Although the data contains information on the full range of physical assets, this paper focuses on two key assets, primary residence and agricultural land. The primacy of land for livelihoods, particularly in developing countries is not debatable where being landless is often a very clear indication of the poor economic status of the household. It is often the last asset to be disposed in times of crises and can make a significant difference to a household's poverty profile (Krishna 2006). The use of housing as a criterion in poverty targeting has focused exclusively on the quality of housing and its associated amenities. Without denying the role of amenities, we contend that home ownership is an equally critical measure of well being, given the vulnerabilities associated with lack of tenure security, particularly for informal settlements in urban spaces. Land and home ownership have to be understood in the larger context of social relations beyond the economic benefits attached to them. Ownership confers status and prestige within one's community and can also be empowering due to potential to control one's immediate environment (Datta, 2006).

A household is deprived if it does not own at least one of these two assets. For the individual measure, in addition to the household indicator, we also include an individual ownership indicator. If the individual is not the owner (individually or joint) of either the house or land then she is considered deprived. The quantity of land owned and the type of house are not taken into consideration under the assumption that having their name on any asset, no matter how small, can be empowering women.

*Empowerment:* Based on field testing, mobility constraints were not found to be relevant for men and therefore these questions were asked only to women in the household. We assume that all adult females in the household have the same level of mobility as the female (primary or secondary) respondent. Since women's empowerment has positive externalities for the household as a whole, a household is assigned the values of its women members for each of the above indicators. At the individual level all men are considered non-deprived in these indicators.

This paper follows the UNDP-MPI methodology in adopting an equal weighting approach. The issue of weighting in multidimensional poverty measures has been much discussed but given that the focus of this study is on demonstrating the usefulness of an individual level poverty measure and not on the robustness of the measure to different weighting schemes per se, we opt for the simplest approach. All four dimensions are weighted equally and within each dimension, all indicators are also given equal weights. The household and individual weights differ due to the variation in the number of indicators within each dimension. For example, at the household level, education receives a weight of 0.25 while each indicator (schooling and child enrolment) are weighted at 0.125. For the individual measure, since the education dimension has only one indicator, it receives a weight of 0.25. Conversely, for the productive assets dimension, the household measure has one indicator (asset ownership) while the individual measure has two indicators (household ownership and individual ownership of assets). Households were evaluated in each indicator based on the indicator cut-offs described in the previous section.

We aggregate the total number of weighted deprivations for each household and individual with the identification of poor based on a poverty cut-off of 30 per cent as per the methodology of the UNDP-MPI. However, we also present a dominance analysis in the appendix which compares results from a range of poverty cut-offs. The results are robust to changes in the poverty cut off specification. For the purposes of this study, an individual or household is poor if they are deprived in 30 per cent or more of the weighted deprivations. For example, an individual deprived in the individual asset ownership indicator (weight of 0.125) and the education indicator (weight of 0.25) would be considered poor since these two deprivations together constitute 37 per cent. We also calculate the average intensity of deprivation among the poor, that is, the average deprivation score among those indentified as poor. The KHAS-MPI is calculated as the product of the head count or the percentage of poor households (or individuals) and the average intensity of deprivation among the poor.

## 5. Results

This section presents the multidimensional poverty measure both at the household and the individual level. After accounting for missing variables, the final sample used in this study is 3,400 households. At the household level, the KHAS-MPI for Karnataka is 0.10 with approximately 25 per cent of the households classified as being multidimensionally poor (Table 2)[2]. On assigning the multidimensional poverty value of their household to individual members, about 22 per cent of all individuals are identified as multidimensionally poor with the poverty rate similar for men (21%) and women (22%). Since men and women within a household have the same deprivation scores, this similarity in the poverty rate seems to suggest that men and women are fairly evenly distributed across poor and non-poor households and that there is no major gender difference in poverty.

**Table 2. Household MPI and poverty rate (poverty cut off = 30%)**

|  | All | Female-headed | Male-headed |
|---|---|---|---|
| KHAS-MPI | 0.10 | 0.09 | 0.10 |
| Average intensity of deprivation among the poor (%) | 40.2 | 41.0 | 40.0 |
| Head count (%) | | | |
| Households | 24.9 | 23.0 | 25.4 |
| Individuals | 21.8 | 20.6 | 22.0 |
| Women | 22.3 | 21.9 | 22.4 |
| Men | 21.2 | 18.4 | 21.7 |
| Total number of households | 3,400 | 699 | 2,701 |

The analysis by male and female-headed households[3] also suggests that the gender differences in poverty are not compelling. Male-headed households show only a slightly higher poverty rate and KHAS-MPI value than female-headed ones. However when individuals are assigned the same poverty score as their households, the poverty rate is slightly higher for women irrespective of headship.

The individual level analysis, however, paints a dramatically different picture. When poverty is evaluated at the individual level (first three columns of Table 3), 49 of individuals are identified as multideminsionally poor which is more than double the poverty head count (22 %) of individuals based on a household level analysis (Table 2). Further, the KHAS-MPI value of 0.232 reflects both a greater intensity of poverty and larger proportions of poor when one considers the individual, not the household, as the unit of analysis. This divergence strongly indicates the presence of large intra-household differences in poverty.

Large gender differences in poverty are also highlighted using the individual level analysis. At 68 per cent, the poverty rate among women is more than double the poverty rate among men (30%) with the consequence that the majority of the poor are women (71%). This steep difference in the poverty rates of men and women is completely masked when poverty is conceptualised at the household level (Table 2). Poor women also experience greater intensity of deprivation on an average (50% in comparison to 42% for men) and therefore have a much higher KHAS-MPI value of 0.335 (compared to 0.123 for men). Since all men are non-deprived in the empowerment dimensions, it is possible that the difference between the poverty rate is biased against women. We therefore estimate the KHAS-MPI without the empowerment dimension while maintaining the equal weighting approach. Thus, the weight of the empowerment dimension is redistributed equally among the remaining three dimensions.

**Table 3. Individual multidimensional poverty index and poverty rate (poverty cut off = 30%)**

|  | All dimensions | | | Without empowerment | | |
|---|---|---|---|---|---|---|
|  | Total | Women | Men | Total | Women | Men |
| Number of poor individuals | 5,476 | 3,890 (71.0%) | 1,586 (29.0%) | 6,221 | 3,690 (59.3%) | 2,531 (40.7%) |
| KHAS-MPI | 0.232 | 0.335 | 0.123 | 0.298 | 0.359 | 0.233 |
| Average intensity of deprivation (%) | 47.7 | 50.0 | 41.9 | 52.6 | 54.8 | 48.3 |
| Head count (%) | 49.4 | 68.3 | 29.5 | 56.2 | 64.8 | 46.8 |
| Number of individuals | 11,092 | 5,691 | 5,401 | 11,092 | 5,691 | 5,401 |

The individual KHAS-MPI and the poverty headcount rate are both higher when empowerment is excluded. This is likely due to the increased weight on the education indicator. Without empowerment, the remaining three dimensions receive a weight of .33 each. Since there is only one education indicator, anyone deprived in education is automatically multidimensionaly poor which also increases the number of individuals classified as poor. What is notable though, is that despite the change in specification, substantial gender differences in poverty persists. The majority of the poor (59%) are still women, the poverty rate for women continues to be substantially higher (65% compared to 47% for men) and the female KHAS-MPI value continues to be higher than the male KHAS-MPI. The empowerment dimension is clearly not the only contributor to the gender difference in poverty; deprivations in the other dimensions are just as critical to understanding why and men and women experience poverty differentially. At the same time, empowerment and individual agency constitute an important aspect of multidimensional poverty and its contribution to gender differences should not be ignored. Therefore, all further analyses are based on the KHAS-MPI that includes empowerment.

Two crucial differences emerged in the comparison between the household level and the individual level analysis of poverty. First, the poverty rate is much higher when evaluated at the individual level. Second, a substantial gender differences can be seen at the individual level but not at the household level. This suggests that real differences exist in the intra-household distribution of resources (physical and human capital in this case) as well as in the ability to participate in decisions that are of importance. It also points to gender as an important axis of differentialtion. To further disentangle the factors driving these differences, we examine the the deprivation rates among the individuals and households in each of the indicators. For ease of exposition, individual poor refers to those who have been classified as such based on the individual level analysis.

Among households, the deprivation rates in the empowerment dimension is high as also in access to basic amenities (Table 4). Deprivation in education even among poor households is low which is in sharp contrast to the experience of individuals. Substantial improvements in school enrollment rates in Karnataka have been reported in recent years (ASER 2011). However the enrollment efforts have focused on the young school going age-group. Older adults therefore continue to be deprived in education. Deprivation rates for all men and women are higher than the household rate. Poor men and women in particular experience very large deprivations in education. The deprivation rate in education is largest among poor men (83%). This divergence between the household and the individual helps explain why poverty rate is much higher when evaluated at the individual level.

We also see large differences in the asset ownership indicator. Approximately 87 per cent of poor women are deprived in the individual ownership of productive assets indicator. In comparison, the deprivation rate in the productive assets category for poor households is relatively lower at 49 per cent. The deprivation rate in individual assets for poor men (51%) is also considerably smaller than poor women. Thus, there are substantial differences in the distribution of economic resources among individuals within households.

Table 4. Deprivation rate for the multidimensional poverty indicators

|  | All Households % | All Men % | All Women % | Poor Households (%) | Poor Men (%) | Poor Women (%) |
|---|---|---|---|---|---|---|
| Schooling | 5.8 | 25.9 | 44.2 | 15.4 | 82.7 | 63.5 |
| Child enrollment | 0.8 | -- | -- | 2.2 | -- | -- |
| Consumer durables | 31.3 | 26.8 | 28.5 | 53.1 | 47.4 | 36.3 |
| Floor | 21.1 | 18.6 | 19.5 | 36.3 | 32.9 | 24.6 |
| Water | 11.8 | 11.8 | 13.6 | 9.2 | 9.7 | 12.4 |
| Cooking | 77.4 | 78.5 | 78.4 | 84.8 | 95.5 | 88.1 |
| Sanitation | 62.6 | 61.1 | 60.3 | 81.8 | 87.2 | 71.6 |
| Electricity | 9.6 | 7.4 | 8.3 | 19.4 | 14.9 | 10.9 |
| Household productive asset ownership | 17.3 | 13.9 | 14.4 | 49.0 | 23.0 | 15.7 |
| Individual productive asset ownership | -- | 56.2 | 84.0 | -- | 51.1 | 87.3 |
| Travel to market | 35.2 | 0 | 36.7 | 61.6 | 0 | 50.4 |
| Travel to health facility | 43.2 | 0 | 44.4 | 72.9 | 0 | 59.5 |
| Travel to natal home | 36.8 | 0 | 38.8 | 66.2 | 0 | 51.6 |
| Travel outside village/community/area | 44.1 | 0 | 44.9 | 74.7 | 0 | 60.0 |
| Access health services for own needs | 1.5 | 0 | 1.6 | 2.8 | 0 | 2.0 |
| Total number of observations | 3,400 | 5401 | 5691 | 847 | 1586 | 3890 |
| Average age | -- | 39.1 | 38.1 | -- | 44.0 | 40.4 |

The interesting question to ask then, is, how many poor people actually reside in non-poor households? Almost 65 per cent of poor individuals in fact are from non-poor households (Table 5). Interestingly, when disaggregated by gender, the same pattern holds for both men and women. A greater proportion of poor men and women belong to non-poor rather than poor households. The misclassification is less severe on examining poor households; only 9 per cent of individuals are non-poor in poor households and this figure is dominated by men. From a policy perspective, the first kind of misclassification is worrisome. Many social protection schemes and subsidies target households based on household-level aggregates which would exclude many poor individuals who are present in non-poor households. In fact, poor men and women are more likely to be found in non-poor male headed households (Table 6). These results show the flaws of using the household as the unit in poverty analysis and the use of headship as a proxy for understanding the gendered impact of poverty. Since male-headed households are the majority, more poor individuals whether men or women will in fact tend to be in male-headed households. Therefore intra-household dynamics have an impact on both men and women.

**Table 5. Distribution of poor/non-poor individuals across poor/non-poor households**

| Household level analysis | Individual level analysis (%) | | | | | |
| --- | --- | --- | --- | --- | --- | --- |
| | Poor | | | Non-poor | | |
| | Total | Men | Women | Total | Men | Women |
| Poor | 35.3 | 42.1 | 32.5 | 8.5 | 12.6 | 0.1 |
| Non-poor | 64.7 | 57.9 | 67.5 | 91.5 | 87.4 | 99.9 |
| Number of individual poor | 5,476 | 1,586 | 3,890 | 5,616 | 3,815 | 1,801 |

**Table 6. Distribution of poor individuals across poor and non-poor households, by headship status**

| Household type | Poor men (%) | Poor women (%) |
| --- | --- | --- |
| Male headed | | |
| Poor | 34.6 | 25.6 |
| Non-poor | 48.3 | 54.7 |
| Female headed | | |
| Poor | 7.5 | 6.9 |
| Non-poor | 9.6 | 12.8 |
| Number of individual poor | 1,586 | 3,890 |

We are also able to identify important differences in the contribution of each dimension to poverty among different groups. For poor women in non-poor households, at 36%, education contributes slightly more than one-third to multidimensional poverty (Table 7). Lack of asset ownership (23%) is the second contributor to poverty for this group even though they actually belong to households that do own some assets (contribution of household productive assets indicator is zero). Women in poor households on the other hand are more multidimensionally poor, with all the four dimensions making somewhat equal contribution to their poverty status. Here the household deprivation in productive assets also contributes to the KHAS-MPI.

The differences in the poverty experience of the two groups have important implications for policy. In poor households there is a clear need for more monetary support to improve living standards, and encourage asset accumulation. In non-poor households on the other hand the mere increase in household wealth does not necessarily improve the position of women. Their relative position in terms of bargaining power within the household might even decline if they are shut out of the ownership of assets and if household resources are not allocated to educate women.

**Table 7. Contribution of dimensions/indicators to KHAS-MPI among poor women**

| Dimensions/indicators | Household poverty status | | Marital status[*] | | |
|---|---|---|---|---|---|
| | Poor | Non-poor | Never married | Currently married | Widowed |
| Living standards | 21.2 | 20.5 | 23.0 | 20.5 | 20.2 |
| Education | 23.7 | 36.1 | 12.0 | 31.5 | 44.0 |
| Household productive asset ownership | 10.0 | 0.4 | 5.0 | 4.0 | 3.3 |
| Individual productive asset | 20.4 | 23.4 | 30.0 | 23.4 | 15.1 |
| ownership | | | | | |
| Empowerment | 25.5 | 21.2 | 30.0 | 22.9 | 18.7 |
| Number of individuals | 1,266 | 2,624 | 366 | 2,748 | 699 |

*Deserted women are not included in this category since they might have different circumstances from widows in terms of inheriting the assets of the spouse.

We also examine the poverty experience of poor women based on their marital status as it is one of the factors that affect women's position and access to resources within a household. In comparison to married and single women, poverty among widowed women is less a factor of lack of empowerment or lack of ownership of assets. The biggest influence is lack of education contributing 44% to their poverty score. The poverty of widows is less dependent on lack of asset ownership; not surprising as they are likely to have inherited assets from their spouse and also considered head of their households.

For never married women, on the other hand, education contributes only 12 per cent to their poverty. This is most likely a generational effect. The average age of the women in the never married group is 23 years while for the currently married and widowed groups it is 41 and 60 years, respectively. Education enrollments in Karnataka have largely benefitted the younger women but the gains in education have not translated into greater mobility for these women. The empowerment dimension (along with lack of ownership of assets) is the largest contributor to the multidimensional poverty status of this category of women. These results highlight the importance of not treating women as a homogenous group in poverty analysis.

## 6. Conclusion

This paper constructs an individual level multidimensional poverty measure which highlights important shortcomings in the current poverty discourse. We find that the poverty rate is underestimated when household aggregates are used for analysis; poverty rate calculated using individual-level data is almost double the poverty rate derived from household-level data. This is largely driven by the fact that household resources are not always pooled and used to benefit all members equally. We find that a majority of both poor men and women belong to non-poor households. These individuals would be misclassified as non-poor in a household level poverty analysis. Women, in particular, are completely overlooked in the traditional approach. Current gender analysis relies on using female heads as a proxy for all women while ignoring those who reside in male-headed households. Since several studies including this one has found that female heads are not necessarily worse off than male headed ones, it has resulted in the erroneous conclusion that gender differences in poverty do not exist. However, this study also demonstrates that when poverty is calculated at the individual level, a substantial majority of the poor are women and the poverty experienced by them is more intense in terms of the number of deprivations they face in comparison to men or the average household. In fact women contribute 91 per cent of the total individual KHAS-MPI.

Furthermore, the individual level analysis provides greater support for a multidimensional approach rather than the traditional monetary measures of poverty. Deprivation in the material space as measured through income and consumption are inadequate to describe poverty even for households without accounting for individuals differences. Individual experiences of what it means to live in poverty are varied and based on the specific kinds of deprivations they experience. For example, it is shown that the poverty of poor women in non-poor households is primarily caused by deprivation in education and lack of individual ownership of assets even when the households are mostly non-deprived in ownership of assets.

These differences in the experience of poverty among the diverse groups of poor women highlight the role of individual agency, a crucial component of the capabilities approach to poverty. The ability to control life circumstances can have important implications for individual's ability to avoid chronic deprivations. From a gender perspective, women's empowerment and ability to have a greater voice in household decisions has been shown to have many positive implications for the living conditions of both individuals and households.

These results are important from a policy perspective. Policy makers are moving away from a narrow conceptualization of poverty to a comprehensive understanding of multiple deprivations. However, the fact that households and individuals cannot be equated in poverty analysis is still far from being accepted universally. Certainly, there are conceptual, methodological, and data collection challenges in moving from the household to the individual. But there are compelling reasons to take on such an exercise. Poor individuals in non-poor households would be completely excluded from any policy intervention that targets only poor households. Even within poor households, men and women experience different sets of deprivations. Women are typically deprived in the empowerment, asset ownership and education dimensions. Merely increasing the material wealth of the households where the women reside, therefore will not necessarily translate into fewer deprivations for them. As the household wealth increases, the bargaining position of these women might even decline due to changes in relative wealth positions. If the goal of poverty reduction is a serious consideration, then the assumptions of using household aggregates need revisiting while also attempting to grapple with the complexities of an individual level approach.

Appendix I
**Dominance Analysis**

We calculate gender disaggregated KHAS-MPI for 10 different poverty cutoffs ranging from 10 per cent to 100 per cent deprivation, to see if our result of higher poverty rates among women holds across different poverty lines. Poverty head count is greater for women across the different cutoffs (Figure A1). KHAS-MPI for women dominates KHAS-MPI for men across the different poverty cut-offs (Figure A2). This shows that poverty among women is higher than men irrespective of the deprivation poverty cutoff chosen to define the poor.

Figure A1: Poverty head count by sex for different poverty cutoffs

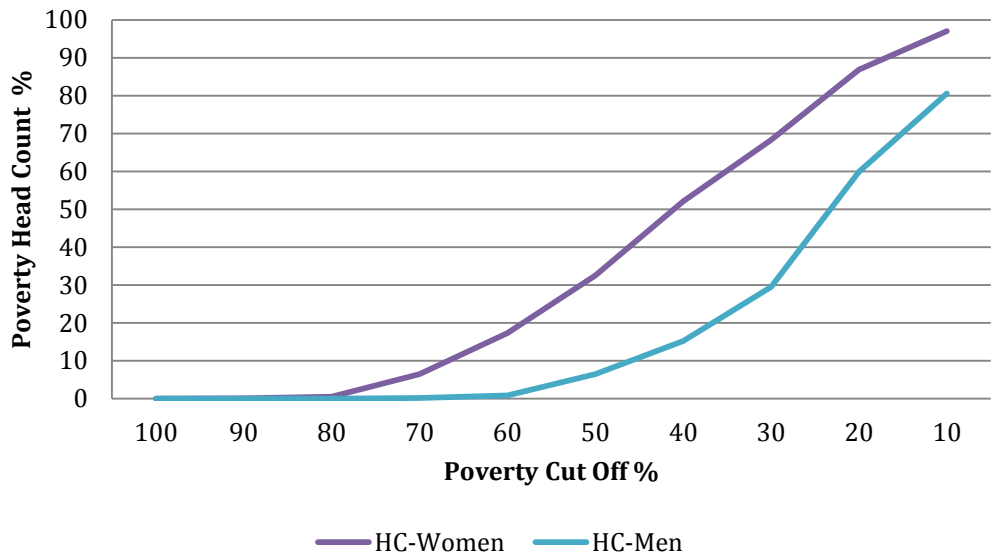

Figure A2: KHAS-MPI by Sex for Different Poverty Cutoffs

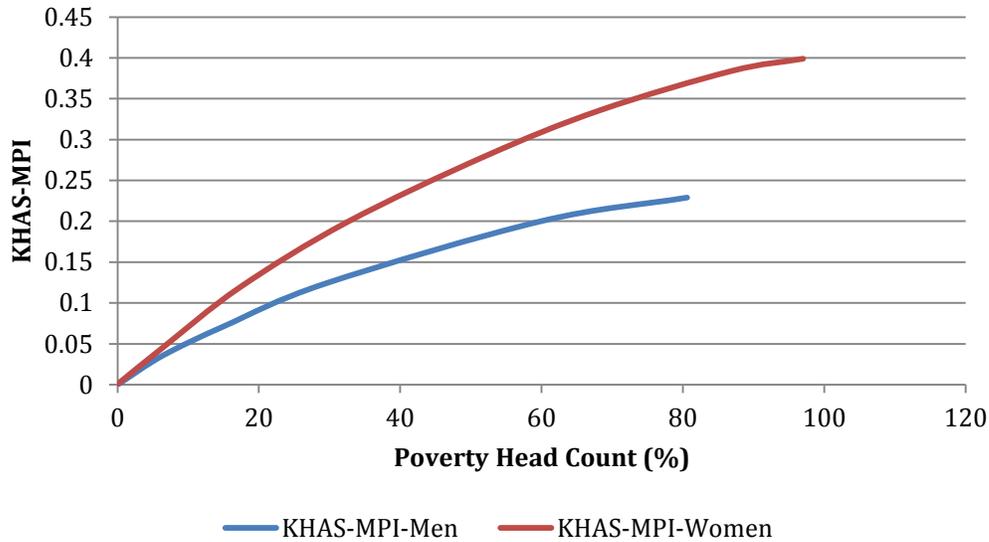

---

[1] The term feminisation of poverty carries several meanings depending on the context and how it is used. This paper focuses on a poverty-related attribute (see Chant (2008) for more details).

[2] We also calculate the household poverty measure excluding the child school enrollment indicator to ensure comparability between the individual and household level analysis. About 26 per cent of households are classified as poor when we exclude child enrollment indicator vs. 25 per cent when we include it. This indicates that including child enrollment in the household measure does not bias the comparison between the household and individual level analysis.

[3] The survey requested households to identify a primary respondent to move away from the traditional headship concept. However, it many households, the primary respondent coincided with the head and thus, is used as a proxy for headship.